# Realizing quantum controlled phase-flip gate through quantum dot in silicon slow-light photonic crystal waveguide

Jie Gao*, Fangwen Sun and Chee Wei Wong

*Optical Nanostructures Laboratory, Columbia University, New York, NY 10027 USA*

**Abstract:** We propose a scheme to realize controlled phase gate between two single photons through a single quantum dot in slow-light silicon photonic crystal waveguide. Enhanced Purcell factor and β-factor lead to high gate fidelity over broadband frequencies compared to cavity-assisted system. The excellent physical integration of this silicon photonic crystal waveguide system provides tremendous potential for large-scale quantum information processing.

* Corresponding author: jg2499@columbia.edu

Atom-cavity system has been well discussed as the critical components for quantum information processing, such as single photon source [1], two-qubit quantum gate operation [2-3] and entanglement generation [4]. Generally optical cavities with high quality factor and small mode volume need to be designed and fabricated to achieve strong coherent interactions between atom and photons. Using a cavity to modify the local density of states (LDOS) is typical limited to a narrow-band spectral region and the photon extraction, scalability and integrability need to be carefully designed. Another possible structure we could use to increase the LDOS could be a slow-light or surface plasmons waveguide rather than a cavity. What can be achieved in high $Q$ cavities such as enhanced emission and strong coupling are also to be expected in the PhC waveguide or nanowire system. Recently guided surface plasmons on conducting nanowires have been studied to achieve strong interaction with individual optical emitters and to create single photon transistors [5]. Signatures of spontaneous emission enhancement of single QD in a PhC waveguide have been demonstrated experimentally [6]. In this Letter, we propose a system which consists of silicon photonic crystal (PhC) waveguides and low dimensional semiconductor quantum dots (QD) for implementing controlled phase-flip (CPF) gate between two flying qubits. In standard PhC waveguide, a single photon can be reflected by a QD in ground state acting as a nearly perfect mirror, which simultaneously gets a $\pi$-phase shift on the reflection. It makes use of tight optical confinement and low group velocity of waveguide modes to influence the emission rates of a localized QD. Enhanced quantum dots emission into photonic crystal waveguide mode provides high gate fidelity over broadband frequencies. Excitation of waveguide mode and extraction of quantum dot emission are extremely efficient in this system and "all integration" is possible.

For standard silicon PhC W1 waveguide, the dispersion diagram of fundamental TE-like propagation mode is shown in Fig. 1(a). A divergent-like LDOS and slow group velocity for wavelengths lying near the PhC waveguide cut-off are expected with a fundamental propagating waveguide mode. Emission of an emitter embedded in a PhC waveguide (at the field maximum of the localized waveguide mode) exhibits a large spontaneous emission enhancement which is proportional to $1/(v_g V_{eff})$. Furthermore, large propagation mode $\beta$ factor (probability of a photon being emitted into a desired waveguide mode regardless of non-radiative decay of the emitter) is obtained throughout the entire propagation spectrum [7]. Fig. 1(b) shows a schematic diagram of the system, with a three-level emitter. Ground and excited states $|g\rangle$ and $|e\rangle$ are coupled via *h*-polarization photons (corresponds to waveguide TE mode) with frequency $\omega_{wg}$. A metastable state $|s\rangle$ is decoupled from waveguide modes but is resonantly coupled to $|e\rangle$ via a classical, optical control field with Rabi frequency $\Omega(t)$. Dynamics of the emitter operator $\sigma_-$ is described by Heisenberg operator equation

$$\frac{d\sigma_-}{dt} = -\frac{1}{2\tau}\sigma_- + i\Delta\omega_{ew}\sigma_- + \kappa a_{in} \qquad (1)$$

where $a_{in}$ ($b_{in}$) is the field operator for the flux of the waveguide input port. The waveguide output fields $a_{out}$ and $b_{out}$ are related to the input fields by $a_{out} = a_{in} + \kappa\sigma_-$ and $b_{out} = b_{in} + \kappa\sigma_-$. $1/\tau$ is the total decay rate of the emitter, and $1/\tau = 1/\tau_{SE} + 1/\tau_{NR} + 1/\tau'$ in which $1/\tau_{SE} = PF/\tau_0$ is the emitter's spontaneous emission rate in PhC waveguide (where *PF* is the Purcell factor), $1/\tau_{NR}$ is the nonradiative decay rate and $1/\tau'$ is related to the spontaneous emission into a continuum of radiation and/or leaky modes. $\Delta\omega_{ew}$ and $\kappa$ are frequency detuning and coupling coefficient between emitter and waveguide mode respectively, and $\kappa = 1/\sqrt{2\tau_{SE}}$.

We use the calculated $\beta$ factor (define as $\beta = \frac{1/\tau_{SE}}{1/\tau_{SE} + 1/\tau'}$) in Ref [7] and consider QD $\tau_0 = 1\text{ns}$ and nonradiative decay could be sub-GHz at low temperature [8], Fig. 1(c) shows the reflectance $R = |r|^2 = |b_{out}/a_{in}|^2$ as a function of normalized frequencies of quantum dots (normalized to $1/\tau_0$) when normalized frequency of waveguide mode(normalized to c/a and c: the vacuum light speed, a: the lattice constant of photonic crystal) f = 0.2662, 0.2668, 0.2682 and 0.2878. It shows that the reflectance curve is getting broader and closer to 1 when the waveguide mode is approaching slower group velocity. The effect of non-radiative emission of quantum dot as a loss mechanism is also shown here in the case of $\tau_{NR} = \tau_0$. The inset shows the reflection coefficient (real part and imaginary part). It indicates that $r \approx -1$ when the QD is on resonance with the waveguide mode with very low group velocity. An input photon is nearly perfectly reflected by the QD, and simultaneously gets a π-phase shift. Similar reflection properties are also shown in surface plasmons nanowire [5], and Ref. [9] describes the ideal waveguide case. The existing losses in this system come from two aspects: one is due to the dephasing process in quantum dot and the other is due to the limited $\beta$ factor.

Based on the QD-PhC waveguide system we described above, we can adopt Duan's protocol in ref. [3] to implement a CPF gate by injecting two photons one by the other for several times and together with single qubit rotations. Here we consider a more compact schematic setup shown in Fig. 2 to realize a CPF gate between two input photon A (target qubit) and B (control qubit). Generally the input photon A or B can be described as $(|h\rangle + |v\rangle)/\sqrt{2}$, and after polarization beam splitter (PBS) only $|h\rangle_A$ or $|h\rangle_B$ enters at the 50:50 beam splitter (BS) and couples into the PhC waveguide from both sides simultaneously. In PhC waveguide, the single photon state is a superposition of the left- and right- propagation waveguide mode. After traveling through the sagnac ring, the photon recombines at the BS and comes out from the same port it entered. Moreover, using the 50:50 BS transformation matrix [10], it is known that photon from one of the input port will gain a π-phase change when it leaves the sagnac ring at the same port. We denote this effect as $|h\rangle_A \xrightarrow{\text{sagnac}} |h\rangle_A, |h\rangle_B \xrightarrow{\text{sagnac}} -|h\rangle_B$. All these free space light paths can also be integrated onto a single chip and we design that the optical paths for $|h\rangle$ and $|v\rangle$ components to be identical. Therefore, when the emitter is on ground state, $|h\rangle_A$ coming from port A will get a π-phase shift after reflecting by QD and leaves our system as $-|h\rangle_A$. The implementation of CPF gate between photon A and B consists of three steps, and first we show initial and final states of the QD-photon system in the following description to illustrate the states evolution after each step in ideal case:

(I) First we initialize the emitter in ground state and apply a control field $\Omega(t)$ simultaneous with the arrival of single photon B. The control field (properly choose to be impedance matched [11]) will result in capture of the incoming single photon (using $|vac\rangle_B$ to describe *h*-polarized B photon after storage) while inducing QD state flips from $|g\rangle$ to $|s\rangle$.

$$|h\rangle_B|g\rangle_{QD} \to |vac\rangle_B|s\rangle_{QD}, |v\rangle_B|g\rangle_{QD} \to |v\rangle_B|g\rangle_{QD} \tag{2}$$

(II) Next we send photon A into the system at this time. Only when emitter is on ground state $|g\rangle$, QD-waveguide system will reflect photon $|h\rangle_A$ and introduce a π-phase shift on this photon simultaneously. The reflected photon will finally leave from port A as below:

$$|h\rangle_A|vac\rangle_B|s\rangle_{QD} \to |h\rangle_A|vac\rangle_B|s\rangle_{QD}, |v\rangle_A|vac\rangle_B|s\rangle_{QD} \to |v\rangle_A|vac\rangle_B|s\rangle_{QD},$$

$$|h\rangle_A|v\rangle_B|g\rangle_{QD} \to -|h\rangle_A|v\rangle_B|g\rangle_{QD}, |v\rangle_A|v\rangle_B|g\rangle_{QD} \to |v\rangle_A|v\rangle_B|g\rangle_{QD} \tag{3}$$

(III) Finally we can choose the same $\Omega(t)$ to drive the emitter from $|s\rangle$ back to $|g\rangle$, and retrieve single photon $|h\rangle_B$ as a time reversal process of (I). The retrieval process can be expressed as $|vac\rangle_B|s\rangle_{QD} \rightarrow |h\rangle_B|g\rangle_{QD}$. The retrieval photon generated in PhC waveguide is exactly the same as the input photon in (I), but remember later it will get a π-phase change when it leaves the BS.

$$\begin{aligned} &|h\rangle_A|vac\rangle_B|s\rangle_{QD} \rightarrow -|h\rangle_A|h\rangle_B|g\rangle_{QD}, |v\rangle_A|vac\rangle_B|s\rangle_{QD} \rightarrow -|v\rangle_A|h\rangle_B|g\rangle_{QD}, \\ &-|h\rangle_A|v\rangle_B|g\rangle_{QD} \rightarrow -|h\rangle_A|v\rangle_B|g\rangle_{QD}, |v\rangle_A|v\rangle_B|g\rangle_{QD} \rightarrow |v\rangle_A|v\rangle_B|g\rangle_{QD} \end{aligned} \quad (4)$$

After these three steps, we have achieved

$$\begin{aligned} &|\varphi\rangle_{initial} = |h\rangle_A|h_B\rangle + |v\rangle_A|h\rangle_B + |h\rangle_A|v\rangle_B + |v\rangle_A|v\rangle_B \\ \Rightarrow \quad &|\varphi\rangle_{ideal} = -|h\rangle_A|h\rangle_B - |v\rangle_A|h\rangle_B - |h\rangle_A|v\rangle_B + |v\rangle_A|v\rangle_B \end{aligned} \quad (5)$$

This ideal photon states evolution shows the successful implementation of controlled phase flip gate operation, which preserves the final phase of A and B photons relative to input only when they are both in *v* polarization, otherwise the final phase will get a π-phase change. The emitter will go back to original ground state after the gate operation.

Now we consider non-ideal cases which include detuning between emitter and waveguide, nonradiative decay of the emitter and experimentally achievable values of low group velocities, as well as photon storage and retrieval efficiency. We include all the above loss mechanisms and experimental limitations into the photon loss during the gate operation in Fig. 3(a). Not surprisingly the photon loss decreases to very low level when we operate at the slow-light PhC waveguide frequencies and the photon loss increases when the non-radiative decay of the quantum dot is comparable to the radiative decay. Gate fidelity of the CPF gate which describe the difference between the real output photon state and the ideal case in eqn.(5) is an important measurement of the quality of the scheme. We note that the slow-light PhC waveguide mode propagation loss and the insertion loss (when coupling and extracting light from PhC waveguide) do not decrease the gate fidelity. In step (I) and (III), we need coherent storage and retrieval of a single photon $|h\rangle_B$ and the store/retrieval efficiency degrades the gate fidelity. Reversible transfer of coherent light to and from the internal state of a single trapped atom in a cavity has already been demonstrated in experiment already [13] and the efficiency could improve up to 90%. In our case, the optimal storage strategy is splitting the incoming pulse and having it incident from both sides of the emitter simultaneously, which is the time reversal process of a single photon generation. There is a one-to-one correspondence between the incoming pulse shape and the optimal field $\Omega(t)$.The retrieval process in (III) is time-reversal process of the storage process in (I) and both efficiencies are determined by the calculated reflection/transmission coefficient [5,12]. Fig. 3(b) show the QD-waveguide CPF gate fidelity as a function of normalized QD frequency detuning for the gate operation at different waveguide mode frequencies or with different quantum dot. The quick drop of the gate fidelity indicates that we always need tune the input photon frequency to be on resonance with the quantum dot transition, although the on-resonance case suffers more photon loss than the off-resonance case in Fig. 3(a). This frequency match here is not that difficult because of the broad spectral range (~10 *THz* from Fig. 1(a)) of the propagating PhC waveguide mode. Fig. 3(c) show the QD-waveguide CPF gate fidelity as a function of PhC waveguide mode frequencies (on resonance with QD). When $f = 0.2662$ , $\nu_g \approx c/154$ have been measured experimentally [14]. Spontaneous emission rate is enhanced by PF = 30 and leads to β factor nearly 0.998 for a QD located at the field antinodes with the same dipole orientation as the mode polarization [7]. The reflectance peak is as high as 0.988 and leads to gate fidelity up to 0.9999 with $\tau_{NR} = 10\tau_0$ used in the simulations. When $f = 0.2827$ , PF tends towards 1 with normal waveguide group velocity but gate fidelity remains above 0.96. Although the Purcell factor is very low in this case, the reason of the high fidelity is that the QD emission into free space or other leaky modes are highly suppressed inside the photonic crystal band gap and we have large β factor all through the waveguide mode spectral range. It indicates as long as the QD transition is within ~ 2 *THz* (15nm) above the waveguide cut-off frequency, our scheme always gets fidelity greater than 0.99 as well as photon loss smaller than 0.18. Even using quantum dot with low quantum yield (assume $\tau_{NR} = \tau_0$), the gate fidelity is higher than 0.9 within ~2 *THz* frequency range because Purcell factor enhanced the QD spontaneous emission rate into the waveguide mode. Combining to contribution both from Purcell factor and large β factor, our QD-PhC waveguide system has a big advantage compared to cavity-assisted schemes by relaxing frequency matching condition (frequency match between the quantum dot transition and the sharp cavity resonance) by ~ two orders of magnitude or more.

In summary we have proposed a new scheme to realize quantum control phase-flip gate between two photons through photon-QD interaction in a photonic crystal waveguide. Strong optical confinement and low group velocity in photonic crystal waveguide contributes to the high gate fidelity (~0.99) over a tremendous broadband region (2*THz*). In our scheme, excitation and extraction can be extremely efficient and chip-scale integration is possible. All these advantages show QD-photonic crystal waveguide system is very promising to be a critical component in quantum information processing.

The authors thank X.D.Yang and J. F. McMillan for helpful discussion. We acknowledge funding support from the National Science Foundation (NSF ECCS 0747787), DARPA MTO YFA, and the New York State Office of Science, Technology and Academic Research.

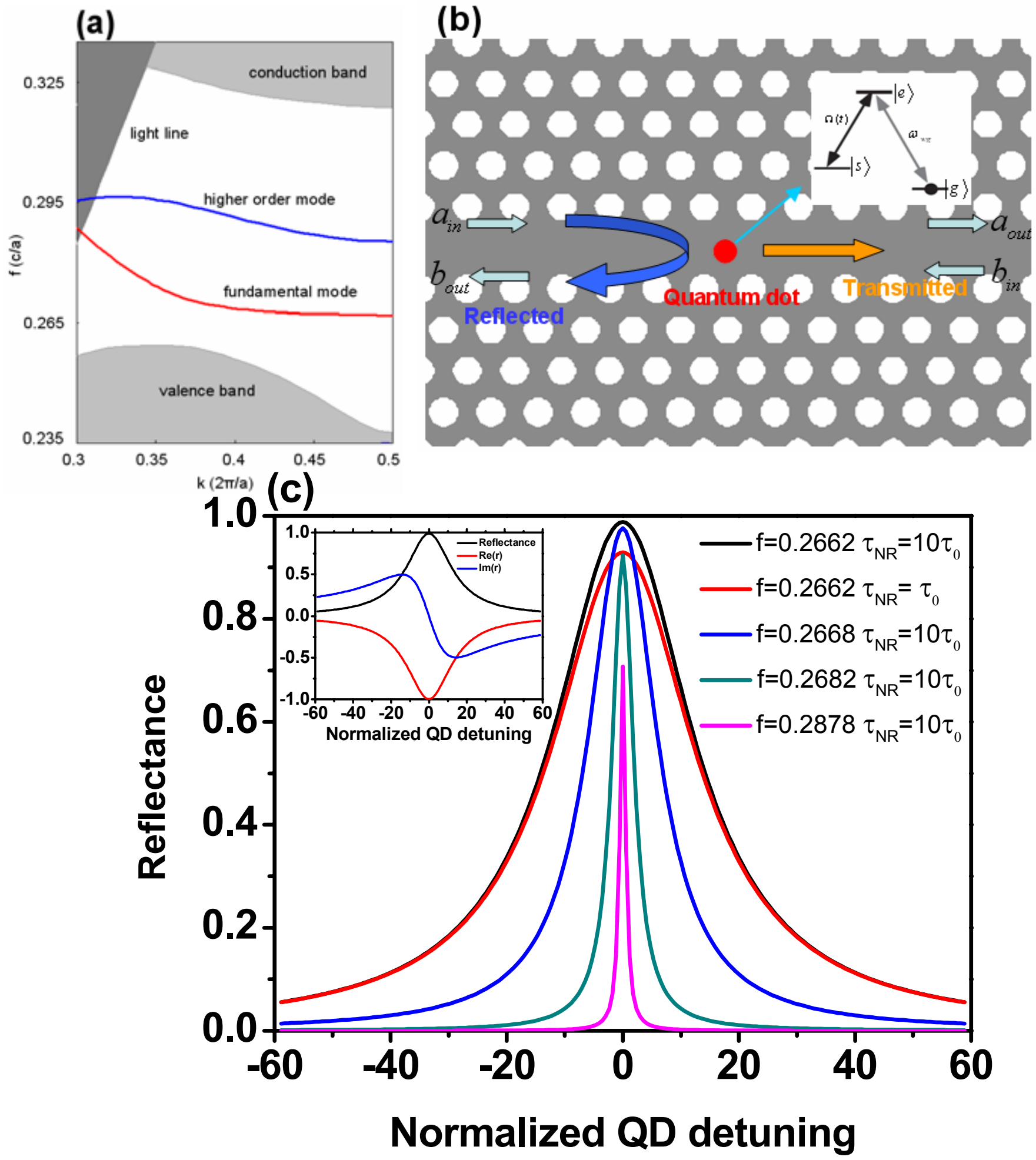

Fig. 1 (a) PhC waveguide band structure within the TE-like band gap. Both fundamental (red) and a higher order mode (blue) are shown. Structure parameters: r = 0.275a, h = 0.5a, ε = 12 and a = 420nm. (b) Schematic diagram of a single incident photon interacts with a near resonant QD. (c) Reflectance as a function of normalized quantum dots detuning frequencies (normalized to $1/\tau_0$) for different normalized PhC waveguide frequencices with quantum dot $\tau_{NR} = 10\tau_0$ (black, blue, green and pink curve) and $\tau_{NR} = \tau_0$ (red curve). Inset: Reflection coefficient (real part and imaginary part) when $f = 0.2662, \tau_{NR} = 10\tau_0$.

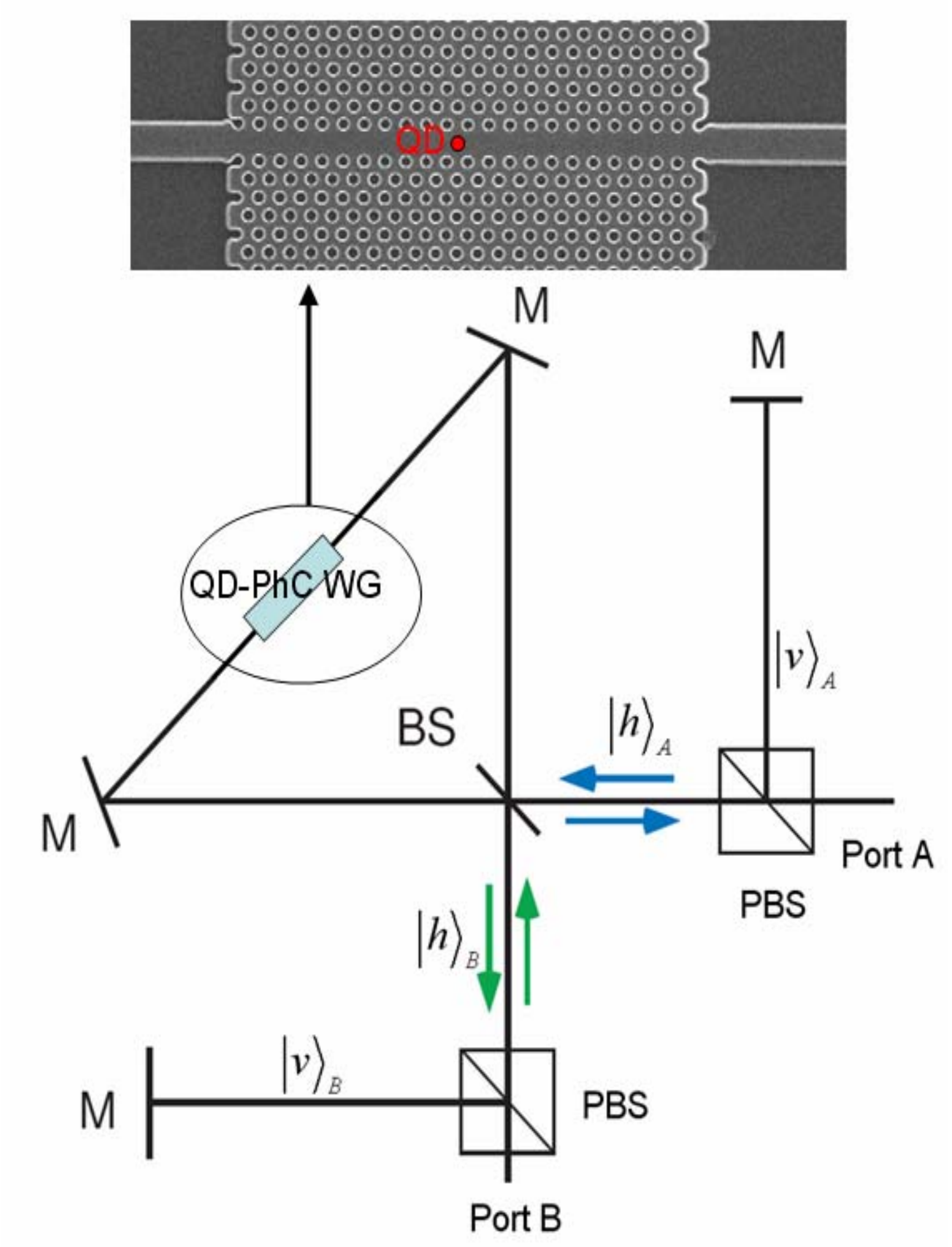

Fig. 2 Schematic setup of CPF gate with sagnac ring. Photon A(B) enters from Port A(B) and the *h*-polarized component (after passing the PBS) interacts with single quantum dot positioned in the PhC waveguide.

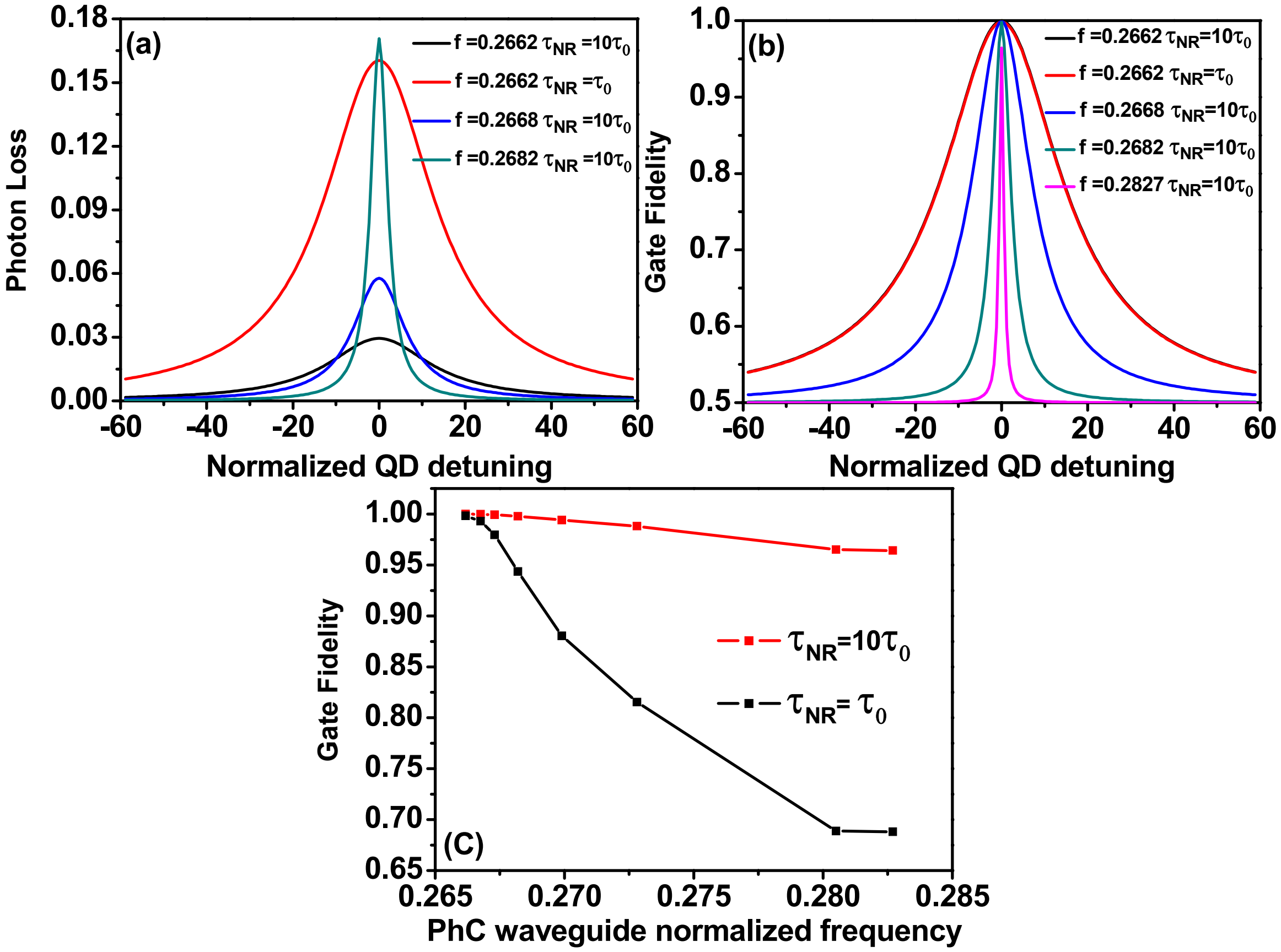

Fig. 3 (a) Photon loss as a function of normalized quantum dot detuning frequencies for CPF gate operated at different PhC waveguide frequencies with quantum dot $\tau_{NR} = 10\tau_0$ (black, blue and green curve) and $\tau_{NR} = \tau_0$ (red curve).(b) gate fidelity as a function of quantum dot detuning frequencies for CPF gate operated at different PhC waveguide frequencies with quantum dot $\tau_{NR} = 10\tau_0$ and $\tau_{NR} = \tau_0$. (c) CPF gate fidelity as a function of normalized PhC waveguide frequencies with quantum dots $\tau_{NR} = 10\tau_0$ (red curve) and $\tau_{NR} = \tau_0$ (black curve).